\documentclass[prl,twocolumn,showpacs,superscriptaddress]{revtex4}
\usepackage{amsmath,amssymb,graphicx,epstopdf,color,bm,soul}

\begin{document}

\title{Stochastic Gross-Pitaevskii Equation for the Dynamical Thermalization of Bose-Einstein Condensates}

\author{I. G. Savenko}
\affiliation{Science Institute, University of Iceland, Dunhagi 3, IS-107, Reykjavik, Iceland}
\affiliation{Division of Physics and Applied Physics, Nanyang Technological University 637371, Singapore}
\author{T. C. H. Liew}
\affiliation{Division of Physics and Applied Physics, Nanyang Technological University 637371, Singapore}
\affiliation{Mediterranean Institute of Fundamental Physics, 31, via Appia Nuova, Roma 00040, Italy}
\author{I. A. Shelykh}
\affiliation{Science Institute, University of Iceland, Dunhagi 3, IS-107, Reykjavik, Iceland}
\affiliation{Division of Physics and Applied Physics, Nanyang Technological University 637371, Singapore}

\begin{abstract}
We present a theory for the description of energy relaxation in a nonequilibrium condensate of bosonic particles. The approach is based on coupling to a thermal bath of other particles (e.g., phonons in a crystal, or noncondensed atoms in a cold atom system), which are treated with a Monte Carlo type approach. Together with a full account of particle-particle interactions, dynamic driving, and particle loss, this offers a complete description of recent experiments in which Bose-Einstein condensates are seen to relax their energy as they propagate in real space and time. As an example, we apply the theory to the solid-state system of microcavity exciton polaritons, in which nonequilibrium effects are particularly prominent.
\end{abstract}

\maketitle


{\it Introduction.---}Bose-Einstein condensates form when multiple
bosonic particles relax their energy, through interaction with other
particles (e.g., phonons), to collect into a low energy state.
Examples in solid-state systems include: condensates of
magnons~\cite{Nikuni1999,Demokritov2007} in quantum spin gases;
indirect excitons~\cite{High2007} in coupled quantum wells; and
exciton-polaritons~\cite{Kasprzak2006,Balili2007,Lai2007} in
semiconductor microcavities.
While Bose-Einstein condensation is conventionally thought of as a
macroscopic occupation of the ground state in thermal equilibrium,
an important issue is that thermal equilibrium in solid state
systems is never perfectly achieved since the particles have finite
lifetimes. The departure from thermal equilibrium is particularly
pronounced in exciton-polariton systems, where a short lifetime of
particles may make the system highly
non-equilibrium~\cite{Wouters2007}. Nevertheless, spontaneous
spatial coherence has been observed~\cite{Kasprzak2006}, although
not necessarily in the ground state of the
system~\cite{Krizhanovskii2009,Maragkou2010}. The theoretical
description of such condensates thus requires a kinetic approach.

In excitonic systems (exciton or exciton-polariton
condensates) scattering with acoustic phonons~\cite{Tassone1997}
offers a mechanism of energy relaxation. It may not offer complete
thermalization, but the effect of energy relaxation is clearly seen
in experiments where a potential gradient is present. The latter can
be created by application of stress~\cite{Balili2007},
optically~\cite{Wertz2010} or structurally
engineered~\cite{Assmann2012}. In this situation, the processes of
energy relaxation and condensate propagation down the gradient are
closely connected. Aside offering a clear demonstration of energy
relaxation, propagating condensates have recently demonstrated
potential as optoelectronic transistors in exciton~\cite{High2007}
and exciton-polariton systems~\cite{Gao2012}.

Theoretically, the dynamics of spatially homogeneous systems have
been described accounting for the exciton-phonon interaction by
means of a system of semiclassical Boltzmann equations
\cite{Porras2002,Haug2005,Kasprzak2008,AmoNature,Cao}. This
approach, however, has serious drawbacks. First, the corresponding
formalism is based on the assumption of full incoherence in the
system, thus quantum states are supposed to be completely
uncorrelated. However, as long as the condensate has been formed,
this is no longer valid. As a consequence, a number of coherent
phenomena such as the onset of superfluidity~\cite{Carusotto2004},
bistability~\cite{Whittaker2005} and hysteresis cannot be described.
Second, the Boltzmann equations can provide us with information
about the occupation numbers in reciprocal ($k$-space) only, whereas
the real space ($x$-space) behavior remains obscure.

On the other hand, if the processes of decoherence are fully
neglected, the state of interacting particles can be treated as a
classical field described by the Gross-Pitaevskii
equation~\cite{Carusotto2004}, which can be modified for incoherent pumping~\cite{Wouters2007,Keeling2008}. Such an approach has been
successful for the description of a variety of recent experiments~\cite{Lagoudakis2011,Christmann2012} in
semiconductor microcavities, including, for example, experiments on
the dynamics of vortices~\cite{Lagoudakis2011}, spatial pattern
formation~\cite{Manni2011,Christmann2012} and spin
textures~\cite{SpinRing,Kammann2012}. However, the Gross-Pitaevskii
equation conserves the particle energy and thus does not account
for phonon-assisted scattering.
Recent models include energy relaxation in a phenomenological way
within a classical stochastic field~\cite{Read2009} or Gross-Pitaevskii type formalism
\cite{Wouters2010}. While these models give
results in agreement with experimental data, they operate with unknown phenomenological
parameters.

In this Letter we introduce a microscopic theory for the description
of energy relaxation in a coherent excitonic ensemble. The
exciton/exciton-polariton field is coupled to a field representing
phonons in the system, which is modeled using stochastic variables.
For simplicity, we consider a resonant coherent injection of
excitons/exciton-polaritons (models of non-resonant excitation,
involving coupling to an exciton reservoir, have been considered
elsewhere~\cite{Wouters2007} and are compatible with our approach).
We consider injection with zero in-plane momentum in a system
containing a potential gradient~\cite{Gao2012}. The potential
gradients accelerate the particles which undergo scattering with
acoustic phonons as they propagate. The latter process leads to the
energy dissipation and thermalization in the system. To give a
complete description of the dynamics we fully account for
exciton-exciton scattering and losses provided by finite lifetime.

We will use parameters corresponding to cavity exciton-polaritons --
the system for which non-equilibrium effects are most clearly
pronounced -- although it should be noted that our formalism is
applicable to the modelling of other systems in which bosons relax
their energy through interactions with an incoherent gas of other
particles. Another example of application of our theory would be the
system of indirect excitons~\cite{High2007}. In the excitonic
optoelectronic transistor (EXOT) \cite{HighScience,Grosso2009},
electric fields introduce potential gradients and the control of the
electric fields modulates the fluxes of excitons. Similar to the
exciton-polariton case, the relaxation down potential gradients
should be mediated by phonons, the process that has not been
addressed theoretically so far to the best of our knowledge.


{\it Theory.---}For simplicity, we consider a 1D system, as in the
case of microwires where there has been a recent experimental focus
on energy relaxation of propagating
condensates~\cite{Wertz2010,Gao2012}. We note however that the
formalism is expected to be compatible with 2D systems as well. We
introduce the quantum field operators for exciton-polaritons,
$\hat{\Psi}_x$ connected with annihilation operators in reciprocal
($k$-) space by the Fourier transforms (${\cal F}$ in short
notation),
\begin{eqnarray}
\nonumber
\hat{a}_k={\cal F}[\hat\Psi_x]=\frac{1}{\sqrt{N}}\sum_x\hat{\Psi}_xe^{-ikx};~~~~~
\hat{\Psi}_x={\cal F}^{-1}[\hat a_k],
\end{eqnarray}
where $N$ is the discretization length.

The Hamiltonian of the system reads
\begin{eqnarray}
\hat{\mathcal{H}}&=&\sum_kE_k\hat{a}^\dagger_k\hat{a}_k
+\sum_x{\cal P}_x\mathrm{e}^{i(k_px-\omega_pt)}\left(\hat{\Psi}^{\dagger}_x+\hat{\Psi}_x\right)\\
\nonumber
&&+\sum_x\left(V_x\hat{\Psi}^\dagger_x\hat{\Psi}_x
+\alpha\hat{\Psi}^\dagger_x\hat{\Psi}^\dagger_x\hat{\Psi}_x\hat{\Psi}_x\right)\\
\nonumber
&&+\sum_{\vec{q}}\hbar\omega_{\vec{q}}\hat{b}^\dagger_{\vec{q}}\hat{b}_{\vec{q}}
+\sum_{\vec{q},k}G_{\vec{q}}\hat{b}_{\vec{q}}\hat{a}^\dagger_{k+q_x}
\hat{a}_k+G_{\vec{q}}^*\hat{b}^\dagger_{\vec{q}}\hat{a}_{k+q_x}\hat{a}^\dagger_k.
%
\end{eqnarray}

The first two lines here correspond to coherent processes. $E_k$ is
the particle dispersion (which is non-parabolic for
exciton-polaritons); the last term in the first line is the coherent
pumping, where ${\cal P}_x$ is the intensity of pump, $k_p$ is the
wavevector of the pump, corresponding to the inclination of an
incident laser beam, and $\hbar\omega_p$ is the pumping energy;
$V_x$ is the potential profile in $x-$space, $\alpha$ is a constant
describing the strength of particle-particle interactions.

The last line in the equation above corresponds to incoherent
processes. To model the interaction with acoustic phonons, we
introduce the Fr\"ohlich
Hamiltonian~\cite{Tassone1997,Hartwell2010}, where the phonons
described by operators $\hat{b}^\dagger_{\vec{q}},\hat{b}_{\vec{q}}$
are considered to be three dimensional. The phonon wavevector is
$\vec{q}=\vec{e}_xq_x+\vec{e}_yq_x+\vec{e}_zq_z$, where $\vec{e}_x$,
$\vec{e}_y$ and $\vec{e}_z$ are unit vectors: $\vec{e}_x$ is in the
wire direction, $\vec{e}_z$ is in the structure growth direction.
The phonon dispersion relation, $\hbar\omega_{\vec{q}}=\hbar
u\sqrt{q_x^2+q_y^2+q_z^2}$, is determined by the speed of sound,
$u$. $G_{\vec{q}}$ is the exciton-phonon interaction strength, whose
calculation can be found elsewhere
\cite{Piermarocchi1996,Carmichael,Hartwell2010,Perlin2008,Yamamoto1999}
(see also the detailed derivation of the formalism in the
``Supplementary Materials'' appended to this Letter).

Using the Heisenberg equations of motion for the operators, one can
write the formal solution for the phonon field as following:
\begin{align}
\hat{b}_{\vec{q}}(t)&=\hat{b}_{\vec{q}}(0)e^{-i\omega_{\vec{q}}t}\notag\\
&-\frac{i}{\hbar}\int_0^tG^*_{\vec{q}}\sum_k\hat{a}_{k+q_x}^\dagger\left(t^\prime\right)\hat{a}_k\left(t^\prime\right)e^{-i\omega_{\vec{q}}\left(t-t^\prime\right)}dt^\prime.
\end{align}

Remembering that phonons represent an incoherent thermal reservoir,
we can replace the term
$\hat{b}_{\vec{q}}(0)e^{-i\omega_{\vec{q}}t}$ by a stochastic
classical variable $b_{\vec{q}}(t)$ (and a similar replacement can
be made for the conjugate field). This represents the analogue of
the Markov approximation within the Langevin approach, when phonons
are assumed to have a randomly varying phase \cite{Carmichael}. The
stochastic variables are complex numbers with real and imaginary
components normalized as follows,
\begin{align}
\left<b_{\vec{q}}^*(t)b_{\vec{q}^\prime}(t^\prime)\right>&=n_{\vec{q}}\delta_{\vec{q}\vec{q}^\prime}\delta(t-t^\prime);\\
\left<b_{\vec{q}}(t)b_{\vec{q}^\prime}(t^\prime)\right>&=\left<b_{\vec{q}}^*(t)b_{\vec{q}^\prime}^*(t^\prime)\right>=0\notag,
\end{align}
where $n_{\vec{q}}$ is the number of phonons in the state with
wavevector $\vec{q}$ determined by the temperature of the system.

The exciton-polariton field dynamics can then be determined solely
by the exciton-polariton operators and the stochastic terms.
Further, within the mean field approximation, the field operator
$\hat{\Psi}_x$ can be treated as a classical variable for condensed
exciton-polaritons, $\psi_x=\langle\hat{\Psi}_x\rangle$ (with the
Fourier image $\psi_k$). Then, physical observables are calculated
over multiple realizations of the evolution dynamics with stochastic
variables $b_k(t)$. The corresponding equation of motion reads (see
the Supplementary Material for the details of the derivation):
\begin{align}
\label{eq:dpsixdt}
i\hbar\frac{d\psi_x}{dt}&={\cal F}^{-1}\left[E_k\psi_k+{\cal S}_k(t)\right]
\notag\\
&+\left[V_x+\alpha\left|\psi_x\right|^2-i\frac{\hbar\gamma}{2}
\right]\psi_x
+{\cal P}_x\mathrm{e}^{ik_px}\mathrm{e}^{-i\omega_pt}
\notag\\
&\hspace{0mm}
+\sum_k\left\{{\cal T}_{-k}(t)+{\cal T}^*_k(t)\right\}\mathrm{e}^{-ikx}
\psi_x,
\end{align}
where we introduced phenomenologically the decay term
$-i\hbar\gamma\psi_x/2$ to account for the radiative decay of
particles~\cite{Carusotto2004}. The constant $\alpha$ describing
polariton-polariton interactions can be estimated as
\cite{Yamamoto1999}: $\alpha\approx E_ba_B^2/(L_y\Delta x)$, where
$L_y$ is the lateral size of the microwire and $\Delta x=L_x/N$ is
the discretization unit.

One can see that interaction with phonons leads to the appearence of
two types of terms. First, one has a term
\begin{equation}
{\cal S}_k(t)=\sum_{q_x}\psi_{k+q_x}(t)\left(\int_0^t{\cal A}_{q_x}(t^\prime){\cal K}_{0}(t-t^\prime)dt^\prime\right)
\label{EqStimulated}
\end{equation}
where ${\cal
A}_{q_x}(t)=\sum_{k^\prime}\psi_{k'+q_x}^\ast\left(t\right)\psi_{k'}\left(t\right)$.
The term is proportional to the cube of the polariton field and does not
directly include a stochastic term. It can be thus interpreted as the
term corresponding to the emission of phonons by a condensate
\emph{stimulated} by polariton density. The convolution integral is
responsible for energy conservation. Note, that the function
\begin{eqnarray}
\nonumber
{\cal K}_{q_x}(t)=-\sum_{q_u,q_z}|G_{\vec{q}}|^2\left(\textrm{e}^{-i\omega_{\vec{q}}t}-\textrm{e}^{i\omega_{\vec{q}}t}\right)\rightarrow\\
\rightarrow 2i\frac{L_z}{2\pi}\frac{a_B}{2\pi}\int\int|G(\vec{q})|^2\textrm{sin}[\omega(\vec{q})t]dq_ydq_z
\end{eqnarray}
is approximately independent of $q_x$ in the range of
$q_x\in(-10^8,10^8)$ $m^{-1}$, and thus in our calculations we put
${\cal K}_{q_x}(t)\approx{\cal K}_0(t)$.

%
%

The stochastic functions ${\cal T}_{q_x}$ and ${\cal T}^*_{q_x}$ in
the last line of Eq.~\eqref{eq:dpsixdt} are defined by the
correlators:

\begin{align}
\left<{\cal T}_{q_x}^*(t){\cal T}_{q_x^\prime}(t^\prime)\right>&=\sum_{q_y,q_y}\left|G_{{q_x,q_y,q_z}}\right|^2n_{q_x,q_y,q_z}\delta_{q_x,q_x^\prime}\delta(t-t^\prime),\notag\\
\left<{\cal T}_{q_x}(t){\cal T}_{q_x^\prime}(t^\prime)\right>&=
\left<{\cal T}_{q_x}^*(t){\cal T}_{q_x^\prime}^*(t^\prime)\right>=0.
\label{EqThermal}
\end{align}
These \emph{thermal} terms contain the phonon field and so are strongly
temperature dependent.  The proportionality of the \textit{thermal}
part to the first power of $\psi_x$ in Eq.~\eqref{eq:dpsixdt} means
that the scattering processes proceed at a rate proportional to the first
power in exciton-polariton density. Consequently, this term corresponds
to the absorbtion of the phonons by the polariton ensemble and their
emission which is stimulated by final state phonon occupancy, but
not exciton-polariton density. The latter processes are instead
represented by the \textit{stimulated} term (Eq.~\ref{EqStimulated})
considered above.


{\it Results.---}We considered an InGaAlAs alloy-based microcavity
and in computations used the following set of parameters: speed of
sound $u=5370$ $m/s$~\cite{Hartwell2010}, $\gamma=1/18$
$ps^{-1}$~\cite{Gao2012}. The function $V_x=V_0-\beta x$ is the
exciton-polariton potential in space, composed of: the potential
defining the walls of the exciton-polariton wire, $V_0$ and a
potential gradient $\beta=9$ $meV/mm$. We consider a linear
potential gradient as in the experiment of Ref.~\cite{Gao2012}. Our
formalism would however be compatible with any potential shape and
could be applied to, for example, the parabolic~\cite{Balili2007} or
staircase~\cite{Assmann2012} potentials studied recently in
experiments. The exciton-polariton dispersion was calculated using a
two oscillator model with cavity photon effective mass
$4\times10^{-5}$ of the free electron mass, Rabi splitting $10$
$meV$ and exciton-photon detuning $2.5$ $meV$ at zero in-plane
wavevector.

\begin{figure}[!b]
\includegraphics[width=0.99\linewidth]{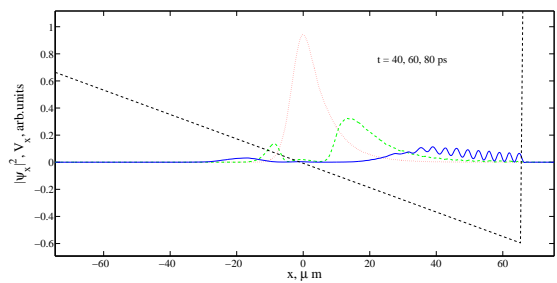}
\caption{Propagation of particles along the potential slope in the
quantum wire due to phonon-assisted relaxation. The curves
correspond to the particle concentration profile in $x$-space for
different times: 40 (red), 60 (green) and 80 ps
(blue). The particles are introduced by a coherent Gaussian pump
of the duration of 20 ps of the theoretical experiment. Particles
propagate along the wire (green and blue curves), suffering losses
caused by their finite lifetime and reflecting from the end of the
wire. The interference of the incoming and reflected waves produces
fringes clearly visible at 80 ps.} \label{FigPsix}
\end{figure}

The main phenomenon we investigated was the relaxation of energy of
exciton-polaritons (thermalization) caused by the interaction with
acoustic phonons. Exciton polaritons were introduced to the system
by localized coherent pump with energy coinciding with the bottom of
lower polariton branch (zero detuning case) which guarantees that we
are outside bistable regime.

\begin{figure}[!b]
\includegraphics[width=0.99\linewidth]{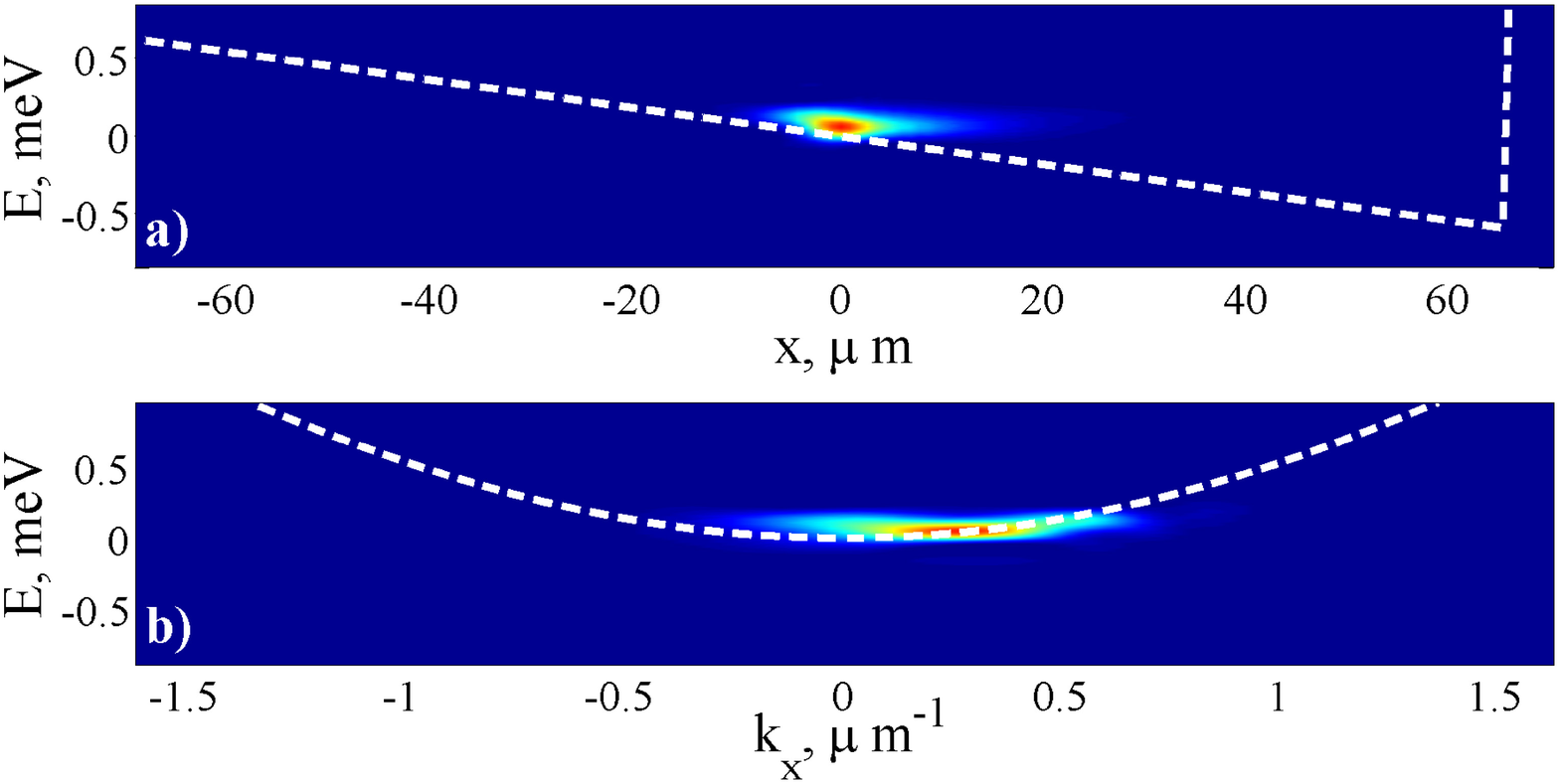}
\includegraphics[width=0.99\linewidth]{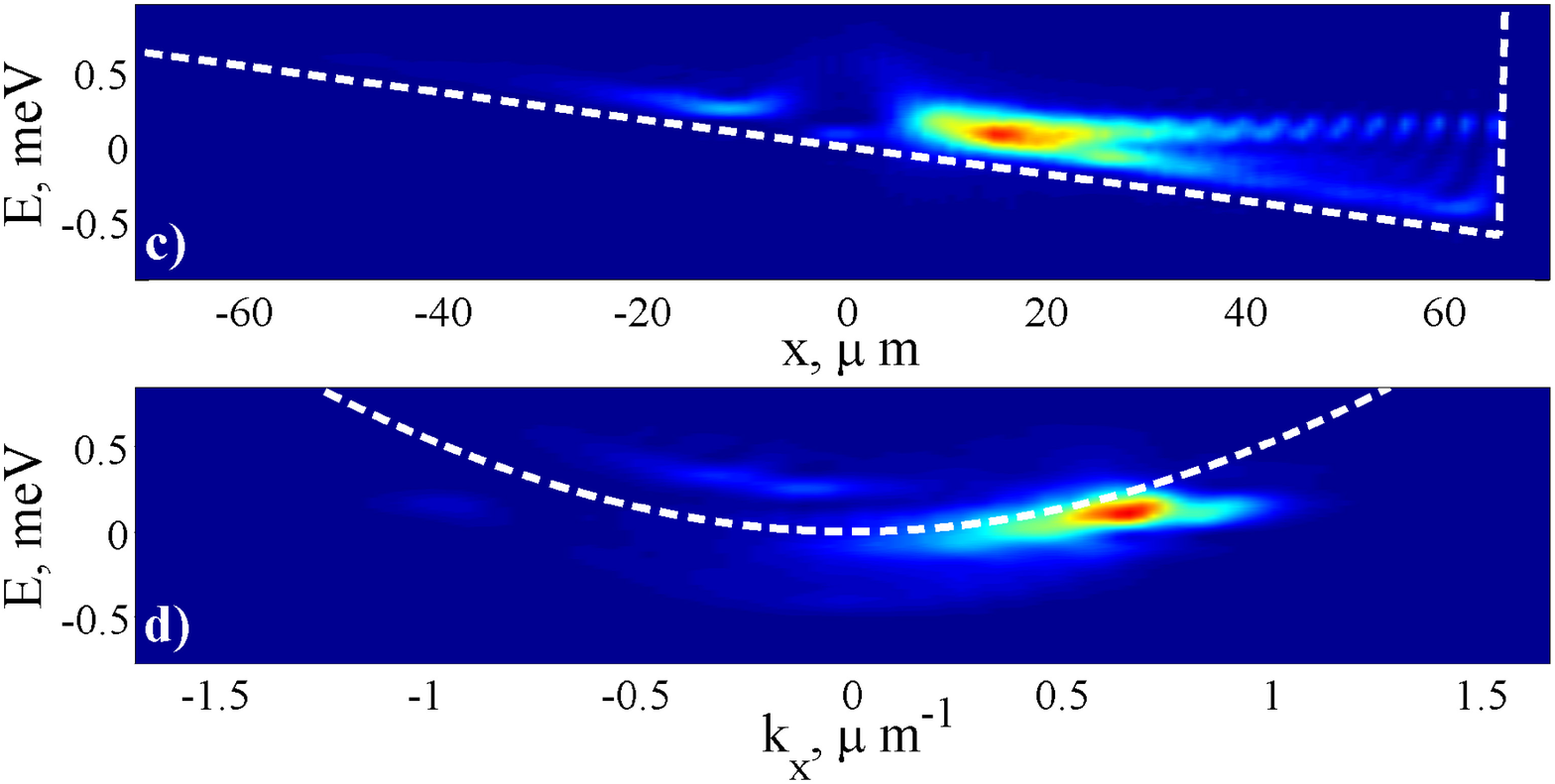}
\includegraphics[width=0.99\linewidth]{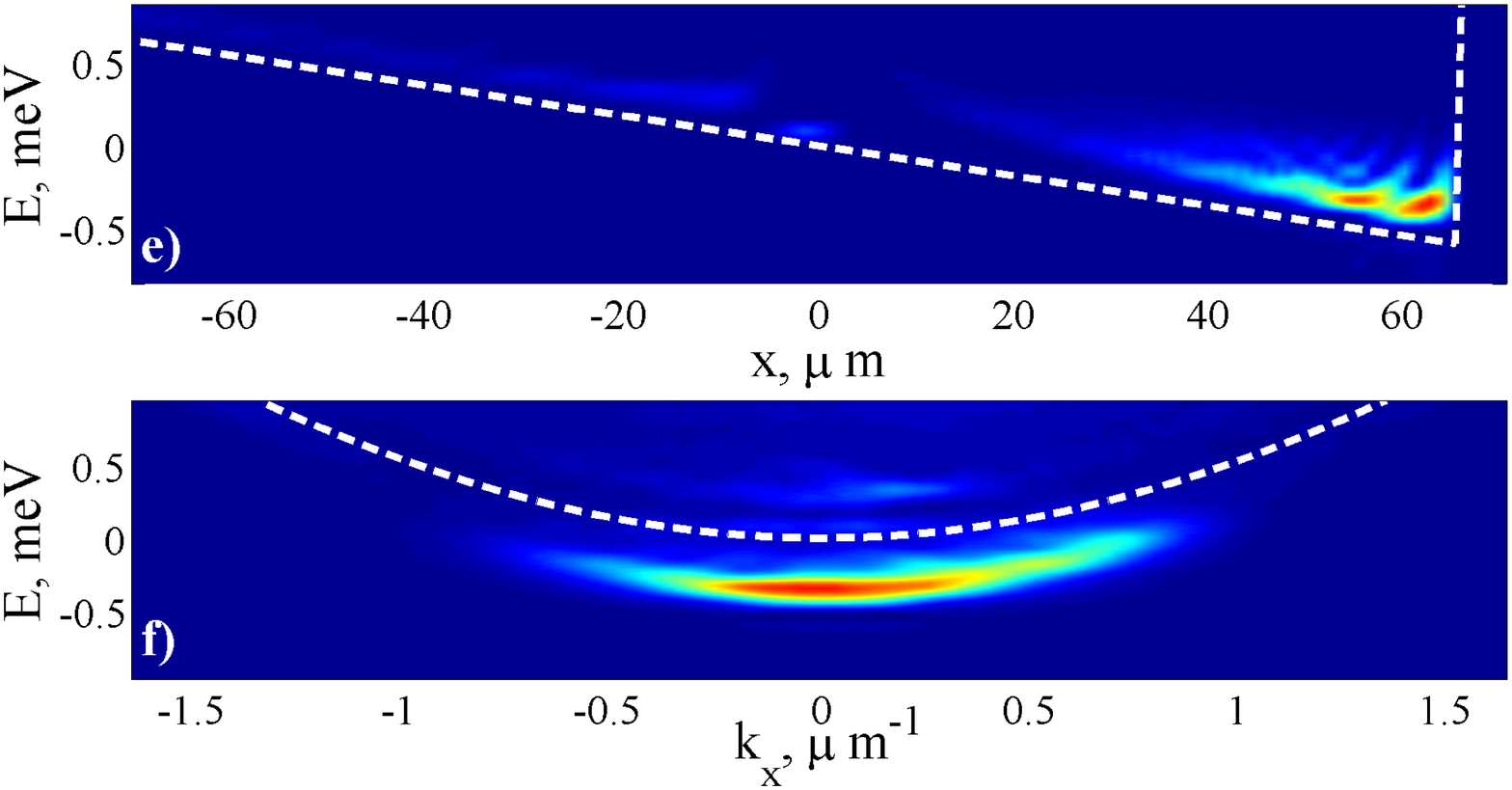}
\caption{Relaxation of energy of exciton-polaritons along the
potential gradient in the regime of $cw$ excitation in a quantum wire
due to phonon-assisted processes in $x$- and $k$-space for different
time ranges: 0-50 ps (a,b), 50-100 ps (c,d) and 100-150 ps (e,f).
White curves show the model potential profile in the real space and
free exciton-polariton dispersion in the reciprocal space used in
calculations. Coherent pumping represents a Gaussian in $x$-space
centered around $x=0$. The inclination angle of the pump is zero,
thus $k_p=0$ and $\hbar\omega_p=E(0)$. Polaritons created at the
pumping spot (a) propagate along the channel (c) and accumulate on
its right hand side (e). The $k$-space behavior shows the decrease
of the conensate energy (b,d,f).} \label{FigExEk}
\end{figure}

Stochastic equation~\eqref{eq:dpsixdt} can be solved
numerically. The results of the modeling are presented in Figs.~\ref{FigPsix}-\ref{FigExEk}a-f.
Figure~\ref{FigPsix} illustrates the propagation of particles
created by a short pulse along the 1D wire. The pumping is switched
on during the first 20 ps of the theoretical experiment, and then it
is off and we observe the decay of intensity due to the finite
particle lifetime. The quantity $|\psi_x|^2$ is depicted for
different times: 20, 40, 60 and 80 ps ($t$ is the parameter). One
can see that the wavepacket of exciton-polaritons propagates along
the channel with time, disperses, and accumulates at the right hand
side. Note, that reflection from the potential jump at the end of
the wire produces pronounced interference fringes clearly visible in
recent experiments \cite{Wertz2012}.

%
%

%
%

Figure~\ref{FigExEk} illustrates the energy relaxation in the system
for three time intervals: 0-50 ps, 50-100 ps and 100-150 ps
correspondingly. It is clearly seen that exciton-polaritons pumped
at $x=0$, $k=0$ (a,b) propagate along the potential slope losing
their energy (c,d) and accumulate on the right-hand side of the wire
(e) thus condensing at $k=0$ in reciprocal space (f). The plots
(a,c,e) snapshot the $x$-space dynamics of the system, the $k$-space
dynamics is depicted in plots (b,d,f). The white dashed curves in
the figures correspond to the potential profile (in $x$-space) and
free polaritons dispersion (in $k$-space). It should be noted, that
the relaxation with phonons is an efficient process, thus particles
rapidly move towards the bottom of the slope and accumulate there;
and their concentration near the pumping spot soon becomes lower
than the concentration in the signal point.


{\it Discussion.---}Comparing with previous theoretical models
aimed at description of energy relaxation in coherent condensates,
we would like to stress that all the parameters, including phonon
scattering rates, are calculated microscopically rather than being
treated in phenomenological way. Furthermore our approach allows the
possibility to treat large regions of direct and reciprocal space
with reasonably low computational cost. This in in contrast to the
theory developed in Ref.~\cite{Wouters2010} where the need to
calculate an evolving particle spectrum in a coarse time window was
extremely memory-demanding. In the approach of Ref.~\cite{Read2009}
the analysis was restricted to a single quantum state and the energy
relaxation introduced phenomenologically. Very recently, in
Ref.~\cite{Galbiati2012}, an approach merging the Boltzmann
equations with a Gross-Pitaevskii treatment was developed where
Boltzmann scattering rates are dynamically calculated from the
mean-field wavefunctions. The theory was successful in the
description of micropillars with a small number of confined states,
but it is not obvious how efficient the theory would be in the
description of many different modes in a large reciprocal space.

Comparing our results to the Monte Carlo approach developed in
Ref.~\cite{Carusotto2005}, it should be noted that we are seeking to
address a different problem. In Ref.~\cite{Carusotto2005} the focus
was on how quantum fluctuations affect the coherence properties of
parametric oscillators. Here the stochastic element describes
fluctuations due to phonons and we are interested in how they can
cause relaxation of propagating condensates. To address coherence
properties, an approach based on a matrix of correlators would be
more appropriate~\cite{Savenko2011,Bozat2012}. Unfortunately, such
an approach is very demanding computationally, requiring matrices of
size $N\times N$, where $N$ represents the number of states in
reciprocal space.

Finally, while we have considered the energy relaxation of
exciton-polaritons via phonon interactions, we should note that
another relaxation mechanism is provided by scattering processes
involving hot excitons with large momentum~\cite{Porras2002}. Such
hot excitons are typically created in non-resonant/incoherently
pumped systems~\cite{Wertz2012} but can be neglected under resonant
coherent excitation which we consider in this manuscript. In
principle, the description of scattering with high momentum excitons
can be accomodated within our formalism. This would allow the
theoretical study of the interplay between both exciton mediated and
phonon mediated scattering processes in extended systems and is an
important direction for future research.


{\it Conclusion.---} We have derived a stochastic Gross-Pitaevskii
equation, where the energy relaxation of bosons is provided by
coupling to an incoherent field, treated as stochastic variable.  As
an example, we applied the theory to the modeling of
exciton-polaritons in a 1D microwire with a potential gradient. The
partial thermalization of exciton-polaritons is observed, together
with their trapping in the real space. This result is of particular
relevance to a variety of recent experiments in exciton and
exciton-polariton systems where the energy relaxation of propagating
Bose-Einstein condensates was reported.

The work was supported by Rannis ``Center of Excellence in
Polaritonics'' and FP7 IRSES POLAPHEN grants. I.G.~Savenko acknowledges
support of the Eimskip foundation.


\begin{widetext}

\section{Supplementary material}
In the present supplemental appendix we derive the stochastic Gross-Pitaevskii-type equation, which accounts for the phonon-mediated relaxation.
In the paper we focus on polariton
propagation in a 1D wire provided by confinement in the lateral
direction, thus the appendix also corresponds to the 1D case. It should be noted, however, that generalization of the
formalism for the 2D case and for any system of interacting bosons is straightforward. We introduce the quantum
field operators for polaritons, $\hat{\Psi}_x$ connected with
annihilation operators in reciprocal ($k$-) space by the Fourier
transforms (${\cal F}$ in short notation),
\begin{eqnarray}
\hat{a}_k&=&{\cal F}[\hat\Psi_x]=\frac{1}{\sqrt{N}}\sum_x\hat{\Psi}_xe^{-ikx};\\
\hat{\Psi}_x&=&{\cal F}^{-1}[\hat a_k]=\frac{1}{\sqrt{N}}\sum_k\hat{a}_ke^{ikx},
\end{eqnarray}
where $N$ is the discretization length.

The Hamiltonian of the system can be represented as a sum of two
parts,
\begin{equation}
\hat{\mathcal{H}}=\hat{\mathcal{H}}_1+\hat{\mathcal{H}}_2.
\end{equation}
The first term $\hat{\mathcal{H}}_1$ corresponds to coherent
processes,
\begin{eqnarray}
\hat{\mathcal{H}}_1&=&\sum_kE_k\hat{a}^\dagger_k\hat{a}_k
+\sum_x\left(V_x\hat{\Psi}^\dagger_x\hat{\Psi}_x
+\alpha\hat{\Psi}^\dagger_x\hat{\Psi}^\dagger_x\hat{\Psi}_x\hat{\Psi}_x\right)
+\sum_x{\cal P}_x\mathrm{e}^{i(k_px-\omega_pt)}\left(\hat{\Psi}^{\dagger}_x+\hat{\Psi}_x\right),
\end{eqnarray}
where $E_k$ is the (non-parabolic) polariton dispersion, $V_x$ is
the potential profile in $x$-space, $\alpha$ is a constant describing the
strength of polariton-polariton interactions. The last sum in the
equation is the coherent pumping term. Here ${\cal P}_x$ is the
intensity of pump, $k_p$ is the wavevector of the pump, corresponding to
the inclination of an incident laser beam, and $\hbar\omega_p$ is the pumping
energy.

To model the interaction with acoustic phonons, we introduce the Fr\"ohlich-type Hamiltonian~\cite{Tassone1997,Hartwell2010}:

\begin{eqnarray}
\label{eq:Hexphon1}
\hat{\mathcal{H}}_2
=\sum_{\vec{q}}\hbar\omega_{\vec{q}}\hat{b}^\dagger_{\vec{q}}\hat{b}_{\vec{q}}
+\sum_{\vec{q},k}
G_{\vec{q}}\hat{b}_{\vec{q}}\hat{a}^\dagger_{k+q_x}
\hat{a}_k+G_{\vec{q}}^*\hat{b}^\dagger_{\vec{q}}\hat{a}_{k+q_x}\hat{a}^\dagger_k
,
%
\end{eqnarray}
where the phonons described by operators
$\hat{b}^\dagger_{\vec{q}},\hat{b}_{\vec{q}}$, unlike polaritons,
are considered to be three dimensional. 
The phonon wavevector is
$\vec{q}=\vec{e}_xq_x+\vec{e}_yq_x+\vec{e}_zq_z$, where $\vec{e}_x$,
$\vec{e}_y$ and $\vec{e}_z$ are unit vectors: $\vec{e}_x$ is in the
wire direction, $\vec{e}_z$ is in the structure growth direction,
and $\vec{e}_y$ is perpendicular to $\vec{e}_x$ and $\vec{e}_z$. The
phonon dispersion relation, $\hbar\omega_{\vec{q}}=\hbar
u\sqrt{q_x^2+q_y^2+q_z^2}$, is determined by the speed of sound,
$u$. 
$G_{\vec{q}}$ is the exciton-phonon interaction strength, 
which can be calculated using the standard electron - deformation potential interaction \cite{Perlin2008,Piermarocchi1996,Yamamoto1999}. 
As a result of elastic scattering of a crystall, there appears an area with relatively higher density and thus higher polarizability. This area attracts an electron efficiently acting as a potential well. 
The relative change in volume $V$ can me expressed through the vector of displacement $\vec{u}_d(\vec{r})$ of a point called the deformation vector:
\begin{equation}
\frac{\delta V}{V}=\nabla\cdot\vec{u}_d.
\end{equation}
Then, the expression for the electron energy can be written in form:
\begin{equation}
E^e[\vec{u}_d(\vec{r})]=E^e_0-d_e\nabla\cdot\vec{u}_d(\vec{r}),
\label{EqDefEn}
\end{equation}
where $E^e_0$ is the energy of electron in a not deformed crystall; the parameter $d_e$ is the constant of the deformation potential for an electron. For the lattice with one atom in an elementary cell, the vector of deformation can be written in form \cite{Perlin2008}:
\begin{equation}
\label{EqDefVec}
\vec{u}_d(\vec{r})=\sqrt{\frac{\hbar}{2\rho VN}}\sum_{\vec{q}}\frac{\zeta_{\vec{q}}}{\sqrt{\omega_{\vec{q}}}}\mathrm{e}^{i\vec{q}\cdot\vec{r}}\left(\hat{b}_{\vec{q}}+\hat{b}^\dagger_{-\vec{q}}\right),
\end{equation}
where $\vec\zeta$ is a real unit vector; $\rho$ is the mean density of the semiconductor material. Putting \eqref{EqDefEn} in \eqref{EqDefVec}, we obtain the scattering strength of an electron on a phonon:
\begin{equation}
G_{\vec{q}}=-id_e\sqrt{\frac{\hbar|\vec{q}|}{2\rho Vu}}=-id_e\sqrt{\frac{\hbar\omega_{\vec{q}}}{2\rho Vu^2}}.
\end{equation}
The same but opposite in sign expression is true for a hole. Combining them and introducing the overlap integrals, we finally come to the expression:
\begin{align}
G_{\vec{q}}=i\sqrt{\frac{\hbar\omega_{\vec{q}}}{2\rho Vu^2}}
\left[
d_eI^e_{\parallel}(q_x,q_y)I^e_{\bot}(q_z)-
d_hI^h_{\parallel}(q_x,q_y)I^h_{\bot}(q_z)\right],
\end{align}
where $d_h$ are the deformation
potentials of the lattice induced by phonons at the points of
location holes. The integrals
$I^{e(h)}_{\parallel}(q_x,q_y)$ and $I^{e(h)}_{\bot}(q_z)$ are the
overlap integrals of the phonon wavefunctions with the electron and
hole wavefunctions, respectively, in the in-plane and growth
directions and can be evaluated following
Ref.~\cite{Piermarocchi1996,Hartwell2010}:
\begin{eqnarray}
I_{\parallel}^{e(h)}(q_x,q_y)&=&\left[1+\left(\frac{m_{h(e)}}{m_e+m_h}\sqrt{q_x^2+q_y^2}a_B\right)^2\right]^{-3/2},\\
I_{\bot}^{e(h)}(q_z)&=&\frac{\pi^2}{\frac{q_zL_z}{2}\left(\pi^2-\left(\frac{q_zL_z}{2}\right)^2\right)}  \textrm{sin}\left(\frac{q_zL_z}{2}\right).
\end{eqnarray}
%
It should be noted that a peculiarity of exciton-polariton systems is the variation of the excitonic fraction of polaritons with the in-plane wavevector, $k$, given by the $k$-dependent Hopfield coefficient. Such a dependence would introduce a $k$-dependence of the non-linear polariton-polariton interaction term and polariton-phonon interaction term. So as not to make our theory too specific to the exciton-polariton case, we neglect such a dependence. This is nevertheless valid in exciton-polariton systems in cases where the excited polaritons have similar in-plane wavevectors. In the excitation scheme that we consider, exciton-polaritons always have a small in-plane wavevector, due to the energy relaxation that limits their kinetic energy. Under such an assumption, the difference in a theory of excitons and of exciton-polaritons lies in the calculation of the non-linear interaction strength, $\alpha$, and of the phonon scattering coefficient, $G_{\vec{q}}$. The nonlinear constant $\alpha$ should be multipled by ${\cal X}^2$ and $G$ by ${\cal X}$, where ${\cal X}\approx 1/2$ for low energy exciton-polaritons and $1$ for excitons.

In cases where one wishes to describe exciton-polaritons with very different in-plane wavevectors, a theory could be based on writing independent equations for the evolution of coupled excitons and photons. In such a case, the interaction of excitons with phonons is of the same form as written in the current theory.


The polariton field dynamics is given by:
\begin{equation}
\frac{d\hat{\Psi}_x}{dt}=
\frac{i}{\hbar}\left[\hat{\mathcal{H}}_1+\hat{\mathcal{H}}_2,\hat{\Psi}_x\right]
,
\label{eq:daxdt}
\end{equation}
where 
the effect of the acoustic phonons is given by the
term
\begin{align}
\label{eq:phononeffect}
\frac{i}{\hbar}\left[\hat{\mathcal{H}}_2,\hat{\Psi}_x\right]=-\frac{i}{\hbar}\sum_{\vec{q}}\left(G_{-q_x,q_y,q_z}\hat{b}_{-q_x,q_y,q_z}+G_{q_x,q_y,q_z}^*\hat{b}^\dagger_{q_x,q_y,q_z}\right)\hat{\Psi}_xe^{-iq_xx}.
\end{align}
The evolution of the acoustic phonons is determined by the equation
\begin{align}
\nonumber
\frac{d\hat{b}_{\vec{q}}}{dt}&=\frac{i}{\hbar}\left[\hat{\mathcal{H}}_2,\hat{b}_{\vec{q}}\right]\\
&=-i\omega_{\vec{q}}\hat{b}_{\vec{q}}-\frac{i}{\hbar}G_{\vec{q}}^*\sum_{k^\prime}\hat{a}_{k'+q_x}^\dagger\hat{a}_{k'}.
\label{eq:PhonField}
\end{align}
The formal solution of Eq.~\eqref{eq:PhonField} reads
\begin{align}
\hat{b}_{\vec{q}}(t)&=\hat{b}_{\vec{q}}(0)e^{-i\omega_{\vec{q}}t}
-\frac{i}{\hbar}G^*_{\vec{q}}\int_0^t\mathrm{e}^{-i\omega_{\vec{q}}\left(t-t^\prime\right)}
{\cal A}_{q_x}(t^\prime)
dt^\prime,
%
\end{align}
where 
\begin{equation}
{\cal A}_{q_x}(t)=\sum_{k^\prime}\hat{a}_{k'+q_x}^\dagger\left(t\right)\hat{a}_{k'}\left(t\right).
\end{equation}
Substitution of this expression into Eq.~\eqref{eq:phononeffect}
gives:
\begin{eqnarray}
\label{eq:daxdt2}
\frac{i}{\hbar}\left[\hat{\mathcal{H}}_2,\hat{\Psi}_x\right]
=&-&\frac{i}{\hbar}\hat{\Psi}_x\sum_{\vec{q}}\left(G_{\vec{q}}\hat{b}_{-q_x,q_y,q_z}(0)\mathrm{e}^{-i\omega_{\vec{q}}t}\right.\left.+G_{\vec{q}}^*\hat{b}_{\vec{q}}^\dagger(0)\mathrm{e}^{i\omega_{\vec{q}}t}\right)\mathrm{e}^{-iq_xx}\\
\nonumber
&-&\frac{1}{\hbar^2}\hat{\Psi}_x\sum_{\vec{q}}\left|G_{\vec{q}}\right|^2\mathrm{e}^{-iq_xx}
\left\{\int_0^t\mathrm{e}^{-i\omega_{\vec{q}}(t-t^\prime)}{\cal A}_{q_x}(t^\prime)dt^\prime\right.
-\left.\int_0^t\mathrm{e}^{i\omega_{\vec{q}}(t-t^\prime)}{\cal A}_{q_x}(t^\prime)dt^\prime\right\},
\end{eqnarray}
where we noted that $G_{-q_x,q_y,q_z}=G_{q_x,q_y,q_z}$ and
$\omega_{-q_x,q_y,q_z}=\omega_{q_x,q_y,q_z}$. Remembering that
phonons represent an incoherent thermal reservoir, we can replace
the terms $\hat{b}_{-q_x,q_y,q_z}(0)e^{-i\omega_{\vec{q}}t}$ and
$\hat{b}_{\vec{q}}^\dagger(0)e^{i\omega_{\vec{q}}t}$ by stochastic
classical variables $b_{-q_x,q_y,q_z}(t)$ and
$b_{q_x,q_y,q_z}^*(t)$, respectively. This approximation is the
analogue of the Markov approximation within the Langevin approach,
when phonons are assumed to have a randomly varying phase. The
stochastic variables are complex numbers with real and imaginary
components drawn from a normal (Gaussian) distribution, normalized
as follows,
\begin{align}
\left<b_{\vec{q}}^*(t)b_{\vec{q}^\prime}(t^\prime)\right>&=n_{\vec{q}}\delta_{\vec{q}\vec{q}^\prime}\delta(t-t^\prime);\\
\left<b_{\vec{q}}(t)b_{\vec{q}^\prime}(t^\prime)\right>&=\left<b_{\vec{q}}^*(t)b_{\vec{q}^\prime}^*(t^\prime)\right>=0,
\end{align}
where $n_{\vec{q}}$ is the number of phonons in the state with
wavevector $\vec{q}$ determined by the temperature of the system.
The summation of the stochastic terms over $q_y$ and $q_z$ in
Eq.~\eqref{eq:daxdt2} can be made by noting that the sum of two
normally distributed stochastic variables again gives a normally
distributed stochastic variable,
\begin{align}
\sum_{q_y,q_z}G_{{-q_x,q_y,q_z}}b_{-q_x,q_y,q_z}(t)&={\cal T}_{-q_x}(t);\\
\sum_{q_y,q_z}G^*_{{q_x,q_y,q_z}}b^*_{q_x,q_y,q_z}(t)&={\cal T}^*_{q_x}(t),
\end{align}
where ${\cal T}_{q_x}$ and ${\cal T}^*_{q_x}$ represent the
temperature dependent or \textit{thermal} part of polariton
scattering on phonons. They are defined by the correlators:
\begin{align}
\left<{\cal T}_{q_x}^*(t){\cal T}_{q_x^\prime}(t^\prime)\right>&=\sum_{q_y,q_y}\left|G_{{q_x,q_y,q_z}}\right|^2n_{q_x,q_y,q_z}\delta_{q_x,q_x^\prime}\delta(t-t^\prime),\\
\left<{\cal T}_{q_x}(t){\cal T}_{q_x^\prime}(t^\prime)\right>&=
\left<{\cal T}_{q_x}^*(t){\cal T}_{q_x^\prime}^*(t^\prime)\right>=0.
\end{align}

The time integrals in Eq.~\eqref{eq:daxdt2} can be simplified using the fact that the function
\begin{eqnarray}
{\cal K}_{q_x}(t)=-\sum_{q_u,q_z}|G_{\vec{q}}|^2\left(\textrm{e}^{-i\omega_{\vec{q}}t}-\textrm{e}^{i\omega_{\vec{q}}t}\right)&\rightarrow&\\
\nonumber
&\rightarrow& 2i\frac{L_z}{2\pi}\frac{a_B}{2\pi}\int\int|G(\vec{q})|^2\textrm{sin}[\omega(\vec{q})t]dq_ydq_z,
\end{eqnarray}
entering Eq.~\eqref{eq:daxdt}, is approximately independent of $q_x$ in
the range of $q_x\in(-10^8,10^8)$ $m^{-1}$, and thus ${\cal
K}_{q_x}(t)\approx{\cal K}_0(t)$.

Further, within the Monte Carlo approach, the field operator
$\hat{\Psi}_x$ can be treated as a classical variable for condensed
polaritons, $\psi_x=\langle\hat{\Psi}_x\rangle$ (with the Fourier
image $\psi_k$). Then, physical observables are calculated
over multiple realizations of the evolution dynamics with stochastic
variable $b_k(t)$. 
We make a last notation
\begin{equation}
{\cal S}_k(t)=\sum_{q_x}\psi_{k+q_x}(t)\left(\int_0^t{\cal A}_{q_x}(t^\prime){\cal K}_{0}(t-t^\prime)dt^\prime\right),
\end{equation}
 and finally obtain the stochastic Gross-Pitaevskii - type equation:
\begin{align}
\label{eq:dpsixdt}
i\hbar\frac{d\psi_x}{dt}&={\cal F}^{-1}\left[E_k\psi_k+{\cal S}_k(t)\right]+\left[V_x-i\frac{\hbar\gamma}{2}+\alpha\left|\psi_x\right|^2
\right]\psi_x
+P_x\mathrm{e}^{ik_px}\mathrm{e}^{-i\omega_pt}
\\
&\hspace{0mm}
+\sum_k\left\{{\cal T}_{-k}(t)+{\cal T}^*_k(t)\right\}\mathrm{e}^{-ikx}
\psi_x,\notag
\end{align}
where we introduced the decay term $-i\hbar\gamma\psi_x/2$ to account for the radiative decay of particles~\cite{Carusotto2004}. The function ${\cal S}_k(t)$ is identified as the \textit{stimulated} part of the phonon-mediated relaxation, where
the convolution integral is responsible for energy conservation during the density-stimulated phonon-assisted process of energy relaxation.

Equation~\eqref{eq:dpsixdt} is the main result of the Letter. It represents the dissipative GP equation and can be applied to various bosonic systems. 
As an example of application, here we present the figure which illustrates energy relaxation in a parabolic trap (see Fig.~\ref{FigExEkParabolic}). 

%
%

\begin{figure}[!b]
\includegraphics[width=0.99\linewidth]{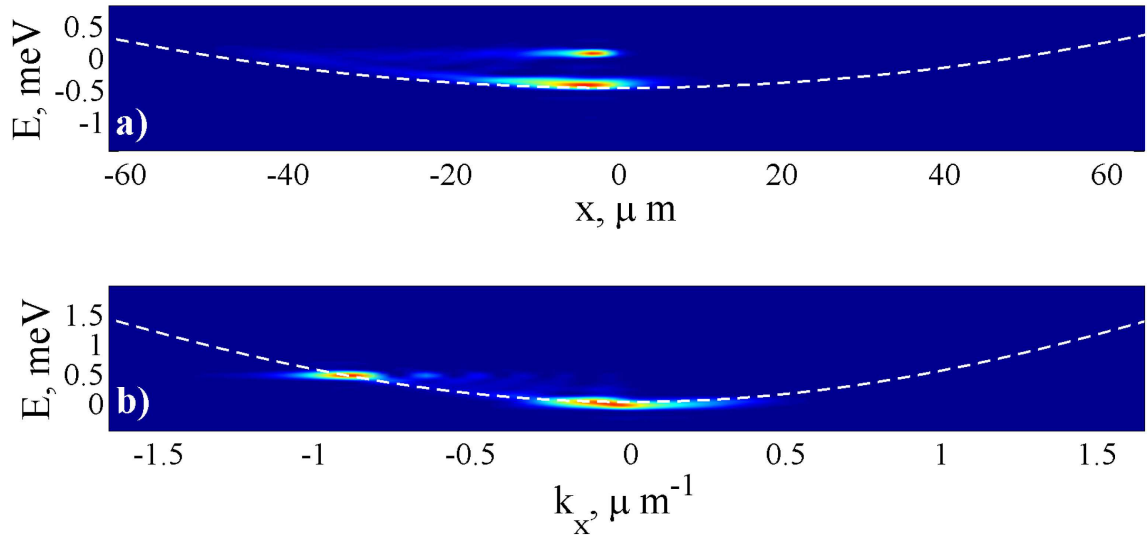}
\caption{Relaxation of energy of exciton-polaritons in the
parabolic trap in the regime of cw excitation in a quantum wire
due to phonon-assisted processes in $x$- and $k$-space for the 
time range: 0-50 ps.
White curves show the Harmonic-oscillator like potential profile in the real space and
free exciton-polariton dispersion in the reciprocal space used in
calculations. Coherent pumping represents a Gaussian in $x$-space
centered around $x=0$ $\mu$m. The inclination angle of the pump in the units of wavevector is $k_p\approx-0.9$ $\mu m^{-1}$, the energy $\hbar\omega_p=0.5$ meV. Polaritons created at the
pumping spot, first, propagate until they meet the potential profile and then along the potential gradient and accumulate in
its bottom (a). The $k$-space behavior shows the decrease
of the conensate energy (b).} \label{FigExEkParabolic}
\end{figure}

%
%

\end{widetext}


\end{document}